\documentclass{article}
\usepackage{spconf,amsmath,graphicx,epsfig,epstopdf}

\title{CN-Celeb: a challenging Chinese speaker recognition dataset}
\name{Y. Fan, J.W. Kang, L.T. Li, K.C. Li, H.L. Chen, S.T. Cheng, P.Y. Zhang, Z.Y. Zhou, Y.Q. Cai, D. Wang}
\address{Center for Speech and Language Technologies, Tsinghua University, China}
\begin{document}
%\ninept
%
\maketitle
\begin{abstract}

Recently, researchers set an ambitious goal of conducting speaker recognition in unconstrained
conditions where the variations on ambient, channel and emotion could be arbitrary.
However, most publicly available datasets are collected under constrained environments, i.e., with
little noise and limited channel variation. These datasets tend to deliver over optimistic performance and
do not meet the request of research on speaker recognition in unconstrained conditions.

In this paper, we present \emph{CN-Celeb}, a large-scale speaker recognition dataset collected `in the wild'.
This dataset contains more than $130,000$ utterances from $1,000$ Chinese celebrities,
and covers $11$ different genres in real world.
Experiments conducted with two state-of-the-art speaker recognition approaches (i-vector and x-vector)
show that the performance on \emph{CN-Celeb} is far inferior to the one obtained on \emph{VoxCeleb},
a widely used speaker recognition dataset.
This result demonstrates that in real-life conditions, the performance of existing techniques might be much worse than it was thought.
Our database is free for researchers and can be downloaded from http://project.cslt.org.

\end{abstract}
\section{Introduction}
\label{sec:intro}

Speaker recognition including identification and verification, aims to recognize claimed identities of speakers.
After decades of research, performance of speaker recognition systems has been vastly improved, and the technique
has been deployed to a wide range of practical applications.
Nevertheless, the present speaker recognition approaches are still far from reliable in unconstrained
conditions where uncertainties within the speech recordings could be arbitrary.
These uncertainties might be caused by multiple factors, including free text, multiple channels,
environmental noises, speaking styles, and physiological status.
These uncertainties make the speaker recognition task highly challenging~\cite{stoll2011finding,zheng2017robustness}.

Researchers have devoted much effort to address the difficulties in unconstrained conditions.
Early methods are based on probabilistic models that treat these uncertainties as an additive Gaussian noise.
JFA~\cite{Kenny07,dehak2011front} and
PLDA~\cite{ioffe2006probabilistic} are the most famous among such models. These models, however, are shallow and
linear, and therefore cannot deal with the complexity of real-life applications. Recent advance in deep
learning methods offers a new opportunity~\cite{hinton2012deep,krizhevsky2012imagenet,simonyan2014very,he2016deep}.
Resorting to the power of deep neural networks (DNNs) in representation learning, these methods
can remove unwanted uncertainties by propagating speech signals through the DNN layer by layer and retain speaker-relevant
features only~\cite{li2017deep}.
Significant improvement in robustness has been achieved by the DNN-based approach~\cite{snyder2018xvector},
which makes it more suitable for applications in unconstrained conditions.

The success of DNN-based methods, however, largely relies on a large amount of data, in particular data
that involve the true complexity in unconstrained conditions.
Unfortunately, most existing datasets for speaker recognition are collected in constrained conditions,
where the acoustic environment, channel and speaking style do not change significantly for each speaker~\cite{fisher1986ther,godfrey1992switchboard,sadjadi20172016}.
These datasets tend to deliver over optimistic performance and
do not meet the request of research on speaker recognition in unconstrained conditions.

To address this shortage in datasets, researchers have started to collect data `in the wild'.
The most successful `wild' dataset may be \emph{VoxCeleb}~\cite{nagrani2017voxceleb,chung2018voxceleb2},
which contains millions of utterances from over thousands of speakers.
The utterances were collected from open-source media using a fully automated pipeline based on computer vision techniques,
in particular face detection, tracking and recognition, plus video-audio synchronization.
The automated pipeline is almost costless, and thus greatly improves the efficiency of data collection.

In this paper, we re-implement the automated pipeline of \emph{VoxCeleb} and collect
a new large-scale speaker dataset, named \emph{CN-Celeb}.
Compared with \emph{VoxCeleb},  \emph{CN-Celeb} has three distinct features:
\begin{itemize}
\item \emph{CN-Celeb} specially focuses on Chinese celebrities, and contains more than $130,000$ utterances
from $1,000$ persons.

\item \emph{CN-Celeb} covers more genres of speech. We intentionally collected data from $11$
genres, including entertainment, interview, singing, play, movie, vlog, live broadcast, speech, drama, recitation and advertisement.
The speech of a particular speaker may be in more than 5 genres.
As a comparison, most of the utterances in \emph{VoxCeleb} were extracted from interview videos. The diversity in genres
makes our database more representative for the true scenarios in unconstrained conditions, but also more challenging.

\item \emph{CN-Celeb} is not fully automated, but involves human check. We found that more complex the genre is, more errors the automated
pipeline tends to produce. Ironically, the error-pron segments could be highly valuable as they tend to be boundary samples.
We therefore choose a two-stage strategy that employs the automated pipeline to perform pre-selection, and then perform human check.
\end{itemize}

%To further compare \emph{CN-Celeb} and \emph{VoxCeleb} in a quantitative way,
%two state-of-the-art speaker recognition baselines (i-vector and x-vector) are built based on them, respectively.
%Experimental results show that both two systems on \emph{CN-Celeb} obtain much inferior performance to that on \emph{VoxCeleb}.
%It demonstrates that this new \emph{CN-Celeb} is a more complex and challenging speaker recognition dataset.

The rest of the paper is organized as follows.
%Section~\ref{sec:related} describes some related work,and
Section~\ref{sec:cnc} presents a detailed description for \emph{CN-Celeb}, and Section~\ref{sec:exp} presents
more quantitative comparisons between \emph{CN-Celeb} and \emph{VoxCeleb} on
the speaker recognition task. Section~\ref{sec:cond} concludes the entire paper.

\section{The CN-Celeb dataset}
\label{sec:cnc}

\subsection{Data description}
\label{sec:data}

The original purpose of the \emph{CN-Celeb} dataset is to investigate the \emph{true} difficulties of speaker recognition
techniques in unconstrained conditions, and provide a resource for researchers to build prototype systems and evaluate
the performance. Ideally, it can be used as a standalone data source, and can be also used with other datasets together, in particular \emph{VoxCeleb}
which is free and large. For this reason, \emph{CN-Celeb} tries to be distinguished from but also complementary to \emph{VoxCeleb}
from the beginning of the design. This leads to three features that we have discussed in the previous section: Chinese focused,
complex genres, and quality guarantee by human check.

In summary, \emph{CN-Celeb} contains over $130,000$ utterances from $1,000$ Chinese celebrities.
It covers $11$ genres and the total amount of speech waveforms
is $274$ hours. Table~\ref{tab:data} gives the data distribution over the genres, and Table~\ref{tab:dur} presents the data distribution over the length of utterances.

\begin{table}[htb!]
 \begin{center}
  \caption{The distribution over genres.}
   \label{tab:data}
     \begin{tabular}{|l|c|c|c|}
       \hline
             Genre            &  \# of Spks  &  \# of Utters     &   \# of Hours   \\
       \hline
             Entertainment    &  483             &   22,064      &   33.67   \\
       \hline
             Interview        &  780             &   59,317      &   135.77  \\
       \hline
             Singing            &  318             &   12,551      &   28.83   \\
       \hline
             Play             &  69              &   4,245       &   4.95    \\
       \hline
             Movie            &  62              &   2,749       &   2.20    \\
       \hline
             Vlog             &  41              &   1,894       &   4.15    \\
       \hline
             Live Broadcast   &  129             &   8,747       &   16.35   \\
       \hline
             Speech           &  122             &   8,401       &   36.22    \\
       \hline
             Drama            &  160             &   7,274       &   6.43    \\
       \hline
             Recitation       &  41              &   2,747       &   4.98    \\
       \hline
             Advertisement    &  17              &   120         &   0.18    \\
       \hline
       \hline
           \textbf{Overall}   &  \textbf{1,000}  & \textbf{130,109}  &  \textbf{273.73}  \\
       \hline
     \end{tabular}
 \end{center}
\end{table}

\begin{table}[htb!]
 \begin{center}
  \caption{The distribution over utterance length.}
   \label{tab:dur}
     \begin{tabular}{|l|c|c|}
       \hline
             Length (s)       &  \# of Utterances    &   Proportion  \\
       \hline
             \textless{2}    &  41,658          &   32.0\%    \\
       \hline
             2-5             &  38,629          &   30.0\%   \\
       \hline
             5-10            &  23,497          &   18.0\%   \\
       \hline
             10-15           &  10,687          &   8.0\%   \\
       \hline
             15-20           &  5,334           &   4.0\%   \\
       \hline
             20-25           &  3,218           &   2.5\%   \\
       \hline
             25-30           &  1,991           &   1.5\%   \\
       \hline
           \textgreater{30}  &  5,095           &   4.0\%   \\
       \hline
     \end{tabular}
 \end{center}
\end{table}

\subsection{Challenges with CN-Celeb}

Table~\ref{tab:comp} summarizes the main difference between \emph{CN-Celeb} and \emph{VoxCeleb}. Compared to
\emph{VoxCeleb}, \emph{CN-Celeb} is a more complex dataset and more challenging for speaker recognition research.
More details of these challenges are as follows.

\begin{itemize}
\item  Most of the utterances involve real-world noise, including ambient noise, background babbling,
music, cheers and laugh.
\item  A certain amount of utterances involve strong and overlapped background speakers, especially in the dram and movie genres.
\item  Most of speakers have different genres of utterances, which results in significant variation in speaking styles.
\item  The utterances of the same speaker may be recorded at different time and with different devices, leading to serious cross-time and cross-channel problems.
\item  Most of the utterances are short, which meets the scenarios of most real applications but leads to unreliable decision.
\end{itemize}

\begin{table}[htb!]
 \begin{center}
  \caption{Comparison between \emph{CN-Celeb} and \emph{VoxCeleb}.}
   \label{tab:comp}
     \begin{tabular}{|l|c|c|}
       \hline
                             &   \emph{CN-Celeb}   &  \emph{VoxCeleb}  \\
       \hline
             Source media    &   bilibili.com      &   youtube.com   \\
       \hline
             Language        &   Chinese           &   Mostly English   \\
       \hline
             Genre           &   11                &   Mostly interview  \\
       \hline
             \# of Spks      &   1,000             &   7,363         \\
       \hline
             \# of Utters    &   130,109           &   1281,762     \\
       \hline
             \# of Hours     &   274               &   2,794     \\
       \hline
             Human Check     &   Yes               &   No        \\
       \hline
     \end{tabular}
 \end{center}
\end{table}

\subsection{Collection pipeline}
\label{sec:pipe}

\emph{CN-Celeb} was collected following a two-stage strategy: firstly we used an automated
pipeline to extract potential segments of the Person of Interest (POI), and then applied a human check
to remove incorrect segments. This process is much faster than purely human-based segmentation,
and reduces errors caused by a purely automated process.

Briefly, the automated pipeline we used is similar to the one used to
collect \emph{VoxCeleb1}~\cite{nagrani2017voxceleb} and \emph{VoxCeleb2}~\cite{chung2018voxceleb2},
though we made some modification to increase efficiency and precision.
Especially, we introduced a new face-speaker double check step that fused the information from both the image and
speech signals to increase the recall rate while maintaining the precision.

The detailed steps of the collection process are summarized as follows.

\begin{itemize}
\item \textbf{STEP 1. POI list design}.
We manually selected $1,000$ Chinese celebrities as our target speakers. These speakers were mostly from the
entertainment sector, such as singers, drama actors/actrees, news reporters, interviewers. Region diversity
was also taken into account so that variation in accent was covered.

\item \textbf{STEP 2. Pictures and videos download}.
Pictures and videos of the $1,000$ POIs were downloaded from the data source (https://www.bilibili.com/) by searching
for the names of the persons. In order to specify that we were searching for POI names, the word `human'
was added in the search queries. The downloaded videos were manually examined and were categorized into the
$11$ genres.

\item \textbf{STEP 3. Face detection and tracking}.
For each POI, we first obtained the portrait of the person. This was achieved by detecting and clipping the face images
from all pictures of that person. The RetinaFace algorithm was used to perform the detection and clipping~\cite{deng2019retinaface}.
Afterwards, video segments that contain the target person were extracted. This was achieved by three steps: (1)
For each frame, detect all the faces appearing in the frame using RetinaFace; (2) Determine if the target person appears
by comparing the POI portrait and the faces detected in the frame. We used the
ArcFace face recognition system~\cite{deng2019arcface} to perform the comparison; (3) Apply the MOSSE face
tracking system~\cite{bolme2010visual} to produce face streams.

\item \textbf{STEP 4. Active speaker verification}.
As in~\cite{nagrani2017voxceleb}, an active speaker verification system was employed to verify if the speech was really spoken
by the target person. This is necessary as it is possible that the target person appears in the video but the speech is from
other persons. We used the SyncNet model~\cite{chung2016out} as in~\cite{nagrani2017voxceleb} to perform the task. This model
was trained to detect if a stream of mouth movement and a stream of speech are synchronized. In our implementation,
the stream of mouth movement was derived from the face stream produced by the MOSSE system.

\item \textbf{STEP 5. Double check by speaker recognition}.

Although SyncNet worked well for videos in simple genres, it failed for videos of complex genres such as movie and vlog.
A possible reason is that the video content of these genres may change dramatically in time,
which leads to unreliable estimation for the stream of the mouth movement, hence unreliable synchronization detection.
In order to improve
the robustness of the active speaker verification in complex genres, we introduced a double check procedure based on speaker
recognition. The idea is simple: whenever the speaker recognition system states a \emph{very low} confidence for the target speaker,
the segment will be discarded even if the confidence from SyncNet is high; vice versa, if the speaker recognition system states
a \emph{very high} confidence, the segment will be retained. We used an off-the-shelf speaker recognition system~\cite{xie2019utterance} to perform
this double check. In our study, this double check improved the recall rate by $30$\% absolutely.

%\item \textbf{STEP 6. Time unions}.
%To deal with the problems (too few volume of videos and too many fragmented pieces) as mentioned in Stage 5,
%we calculate the union of POI speakeing time between Stage 4 and Stage 5.
%Experimental results show that this method can greatly improve the precision-recall rate, and
%the recall rate can be absolutely increased by 30\% among different conditions.

\item \textbf{STEP 6. Human check}.

The segments produced by the above automated pipeline were finally checked by human. According to our experience, this human check
is rather efficient: one could check $1$ hour of speech in $1$ hour.
As a comparison, if we do not apply the automated pre-selection, checking $1$ hour of speech requires $4$ hours.

\end{itemize}

\section{Experiments on speaker recognition}
\label{sec:exp}

In this section, we present a series of experiments on speaker recognition using \emph{VoxCeleb} and \emph{CN-Celeb}, to compare the
complexity of the two datasets.

\subsection{Data}

\emph{VoxCeleb}: The entire dataset involves two parts: \emph{VoxCeleb1} and \emph{VoxCeleb2}. We used
\emph{SITW}~\cite{mclaren2016speakers}, a subset of \emph{VoxCeleb1} as the evaluation set. The rest of \emph{VoxCeleb1}
was merged with \emph{VoxCeleb2} to form the training set (simply denoted by \emph{VoxCeleb}). The training set involves $1,236,567$ utterances
from $7,185$ speakers, and the evaluation set involves $6,445$ utterances from $299$ speakers
(precisely, this is the \emph{Eval. Core} set within \emph{SITW}).

\emph{CN-Celeb}: The entire dataset was split into two parts: the first part \emph{CN-Celeb(T)}
involves $111,260$ utterances from $800$ speakers and was used as the training set;
the second part \emph{CN-Celeb(E)} involves $18,849$ utterances from $200$ speakers and was used as the evaluation set.

\subsection{Settings}

Two state-of-the-art baseline systems were built following the Kaldi SITW recipe~\cite{povey2011kaldi}:
an i-vector system~\cite{dehak2011front} and an x-vector system~\cite{snyder2018xvector}.

For the i-vector system, the acoustic feature involved $24$-dimensional MFCCs plus the log energy,
augmented by the first- and second-order derivatives. We also applied the cepstral mean normalization (CMN)
and the energy-based voice active detection (VAD).
The universal background model (UBM) consisted of $2,048$ Gaussian components, and the dimensionality of the i-vector space was $400$.
LDA was applied to reduce the dimensionality of the i-vectors to $150$. The PLDA model was used for scoring~\cite{ioffe2006probabilistic}.

For the x-vector system, the feature-learning component was a 5-layer time-delay neural network (TDNN).
The slicing parameters for the five time-delay layers were:
\{$t$-$2$, $t$-$1$, $t$, $t$+$1$, $t$+$2$\}, \{$t$-$2$, $t$, $t$+$2$\}, \{$t$-$3$, $t$, $t$+$3$\}, \{$t$\}, \{$t$\}.
The statistic pooling layer computed the mean and standard deviation of the frame-level features from a speech segment.
The size of the output layer was consistent with the number of speakers in the training set.
Once trained, the activations of the penultimate hidden layer were read out as x-vectors.
In our experiments, the dimension of the x-vectors trained on \emph{VoxCeleb} was set to $512$,
while for \emph{CN-Celeb}, it was set to $256$, considering the less number of speakers in the training set.
Afterwards, the x-vectors were projected to $150$-dimensional vectors by LDA, and finally the PLDA model was employed to score the trials.
Refer to~\cite{snyder2018xvector} for more details.

\subsection{Basic results}

We first present the basic results evaluated on \emph{SITW} and \emph{CN-Celeb(E)}.
Both the front-end (i-vector or x-vector models) and back-end (LDA-PLDA) models were trained with the \emph{VoxCeleb} training set.
Note that for \emph{SITW}, the averaged length of the utterances is more than $80$ seconds,
while this number is about $8$ seconds for \emph{CN-Celeb(E)}.
For a better comparison, we resegmented the data of \emph{SITW} and created a new dataset denoted by \emph{SITW(S)},
where the averaged lengths of the enrollment and test utterances are $28$ and $8$ seconds, respectively.
These numbers are similar to the statistics of \emph{CN-Celeb(E)}.

The results in terms of the equal error rate (EER) are reported in Table~\ref{tab:basic}.
It can be observed that for both the i-vector system and the x-vector system, the performance
on \emph{CN-Celeb(E)} is much worse than the performance on \emph{SITW} and \emph{SITW(S)}.
This indicates that there is big difference between these two datasets. From another perspective,
it demonstrates that the model trained with \emph{VoxCeleb} does not generalize well,
although it has achieved reasonable performance on data from a similar source (\emph{SITW}).

\begin{table}[htb!]
 \begin{center}
  \caption{EER(\%) results of the i-vector and x-vector systems trained on VoxCeleb and evaluated on three evaluation sets.}
   \label{tab:basic}
    \scalebox{0.85}{
     \begin{tabular}{|c|c|c|c|c|c|}
       \hline
                      &  \multicolumn{2}{|c|}{Training Set}  &   \multicolumn{3}{|c|}{Evaluation Set}  \\
       \hline
             System    &  Front-end       &  Back-end       &   SITW    &    SITW(S)  &  CN-Celeb(E)    \\
       \hline
       \hline
             i-vector   &  VoxCeleb        &  VoxCeleb       &   5.30    &    7.30     &  19.05        \\
       \hline
             x-vector   &  VoxCeleb        &  VoxCeleb       &   3.75    &    4.78     &  15.52        \\
       \hline
     \end{tabular}
   }
 \end{center}
\end{table}

\subsection{Further comparison}

To further compare \emph{CN-Celeb} and \emph{VoxCeleb} in a quantitative way,
we built systems based on \emph{CN-Celeb} and \emph{VoxCeleb}, respectively.
For a fair comparison, we randomly sampled $800$ speakers from \emph{VoxCeleb}
and built a new dataset \emph{VoxCeleb(L)} whose size is comparable
to \emph{CN-Celeb(T)}. This data set was used for back-end (LDA-PLDA) training.

The experimental results are shown in Table~\ref{tab:ivec}. Note that the performance
of all the comparative experiments show the same trend with the i-vector
system and the x-vector system, we therefore only analyze the i-vector results.

Firstly, it can be seen that the system trained purely on \emph{VoxCeleb} obtained good performance
on \emph{SITW(S)} (1st row). This is understandable
as \emph{VoxCeleb} and \emph{SITW(S)} were collected from the same source.
For the pure \emph{CN-Celeb} system (2nd row), although \emph{CN-Celeb(T)} and \emph{CN-Celeb(E)} are
from the same source, the performance is still poor (14.24\%). More importantly,
with re-training the back-end model with \emph{VoxCeleb(L)} (4th row), the performance
on \emph{SITW} becomes better than the same-source result on \emph{CN-Celeb(E)} (11.34\% vs 14.24\%).
All these results reconfirmed the significant difference between the two datasets, and
indicates that \emph{CN-Celeb} is more challenging than \emph{VoxCeleb}.

\begin{table}[htb!]
 \begin{center}
  \caption{EER(\%) results with different data settings.}
   \label{tab:ivec}
    \scalebox{0.90}{
     \begin{tabular}{|c|c|c|c|c|}
       \hline
                       &  \multicolumn{2}{|c|}{Training Set}  &   \multicolumn{2}{|c|}{Evaluation Set}  \\
       \hline
             System    &  Front-end         &  Back-end      &    SITW(S)  &  CN-Celeb(E)    \\
       \hline
       \hline
             i-vector   &  VoxCeleb          &  VoxCeleb(L)   &    8.34     &   17.43    \\
                        &  CN-Celeb(T)       &  CN-Celeb(T)   &    14.87    &   14.24    \\
                        &  VoxCeleb          &  CN-Celeb(T)   &    12.96    &   15.00    \\
                        &  CN-Celeb(T)       &  VoxCeleb(L)   &    11.34    &   15.50    \\
       \hline
       \hline
             x-vector   &  VoxCeleb          &  VoxCeleb(L)   &    5.93     &   13.64     \\
                        &  CN-Celeb(T)       &  CN-Celeb(T)   &    15.23    &   14.78     \\
                        &  VoxCeleb          &  CN-Celeb(T)   &    10.72    &   11.99     \\
                        &  CN-Celeb(T)       &  VoxCeleb(L)   &    12.68    &   15.62     \\
       \hline
     \end{tabular}
   }
 \end{center}
\end{table}

\section{Conclusions}
\label{sec:cond}

We introduced a free dataset \emph{CN-Celeb} for speaker recognition research.
The dataset contains more than $130k$ utterances from $1,000$ Chinese celebrities,
and covers $11$ different genres in real world.
We compared \emph{CN-Celeb} and \emph{VoxCeleb}, a widely used dataset in speaker recognition,
by setting up a series of experiments based on two state-of-the-art speaker recognition
models. Experimental results demonstrated that \emph{CN-Celeb} is significantly different from \emph{VoxCeleb},
and it is more challenging for speaker recognition research. The EER performance we obtained in this paper
suggests that in unconstrained conditions, the performance of the current speaker recognition techniques
might be much worse than it was thought.

%\vfill\pagebreak

% References should be produced using the bibtex program from suitable
% BiBTeX files (here: strings, refs, manuals). The IEEEbib.bst bibliography
% style file from IEEE produces unsorted bibliography list.
% -------------------------------------------------------------------------
\bibliographystyle{IEEEbib}
\bibliography{strings,refs}

\end{document}